%% file: main.tex
\documentclass[amsmath,amssymb,reprint,aps,prl,floatfix,superscriptaddress]{revtex4-2}
\usepackage{bm}
\usepackage{amsmath}
\usepackage{graphicx}
\usepackage{braket}
\usepackage{bbm}
\usepackage[hidelinks]{hyperref}

\begin{document}

\title{Measuring the joint spectral mode of photon pairs using intensity interferometry}

\author{G.S. Thekkadath}
\email{g.thekkadath@imperial.ac.uk}
\affiliation{Department of Physics, Imperial College London, Prince Consort Rd, London SW7 2AZ, UK}
\affiliation{National Research Council of Canada, 100 Sussex Drive, Ottawa, K1A 0R6, Canada}

\author{B.A. Bell}
\affiliation{Department of Physics, Imperial College London, Prince Consort Rd, London SW7 2AZ, UK}

\author{R.B. Patel}
\affiliation{Department of Physics, Imperial College London, Prince Consort Rd, London SW7 2AZ, UK}
\affiliation{Clarendon Laboratory, University of Oxford, Parks Road, Oxford, OX1 3PU, UK}

\author{M.S. Kim}
\affiliation{Department of Physics, Imperial College London, Prince Consort Rd, London SW7 2AZ, UK}

\author{I.A. Walmsley}
\affiliation{Department of Physics, Imperial College London, Prince Consort Rd, London SW7 2AZ, UK}

\begin{abstract}
The ability to manipulate and measure the time-frequency structure of quantum light is useful for information processing and metrology.
Measuring this structure is also important when developing quantum light sources with high modal purity that can interfere with other independent sources.
Here, we present and experimentally demonstrate a scheme based on intensity interferometry to measure the joint spectral mode of photon pairs produced by spontaneous parametric down-conversion.
We observe correlations in the spectral phase of the photons due to chirp in the pump.
We show that our scheme can be combined with stimulated emission tomography to quickly measure their mode using bright classical light.
Our scheme does not require phase stability, nonlinearities, or spectral shaping, and thus is an experimentally simple way of measuring the modal structure of quantum light. 
\end{abstract}

\maketitle

The modal structure of light, such as its spatial and spectral shape, is fundamental to its use in probing and manipulating matter and in extracting information from optical beams. 
For instance, light's time-frequency structure is particularly well-suited for encoding information because it provides a high-dimensional alphabet that is compatible with optical fiber networks~\cite{agrawal2012fiber}.
In quantum optics, nonlinear optical processes are used to generate ultrashort pulsed photon pairs and to prepare quantum states such as single photons~\cite{lvovsky2001quantum,mosley2008heralded}, squeezed vacuum~\cite{slusher1987pulsed}, and entangled frequency combs~\cite{roslund2014wavelength}.
The fast timescale of these quantum states is attractive for quantum information processing~\cite{brecht2015photon,slussarenko2019photonic,fabre2020modes} and metrology~\cite{lamine2008quantum,dorfman2016nonlinear,lyons2018attosecond,ansari2021achieving}, but poses additional challenges for their characterization~\cite{ren2011time,ren2012analysis,gianani2020measuring} and manipulation~\cite{peer2005temporal,eckstein2011quantum,brecht2014demonstration}.
In particular, characterizing their time-frequency structure requires measuring the spectral amplitude and phase of the generated photon pairs~\cite{wasilewski2006pulsed}.
This poses two main challenges.

Firstly, while a photon's spectral amplitude can be measured using a spectrometer, its spectral phase is more challenging to measure.
There are self-referencing solutions to determining the spectral phase of an optical pulse~\cite{walmsley2009characterization}, but these are often implemented using self-induced or externally controlled optical nonlinear devices~\cite{wong1994analysis}.
It is not possible to use the former for single photons because of the very weak electric field strengths, and while the latter approach has been demonstrated~\cite{davis2018measuring,ansari2018tomography,zhang2019delta,ogawa2021direct}, it comes with significant experimental complexity. 
Alternatively, a purely linear solution is possible by interfering the unknown signal pulse with a reference pulse and performing spectrally-resolved detection~\cite{froehly1973time, fittinghoff1996measurement,beck2001joint,thiel2020single}.
This generally requires a reference whose mode is known completely (in both spectral amplitude and phase), and which is phase stable with respect to the signal.
Similar schemes have also been demonstrated without spectrally-resolved detection, but required scanning the reference in order to reconstruct the spectral mode of the signal~\cite{wasilewski2007spectral,polycarpou2012adaptive,medeiros2014full,qin2015complete,tiedau2018quantum}.

The second challenge is that photon pairs can exhibit time-frequency correlations which must be uncovered using joint measurements~\cite{franson1989bell}.
This challenge has been partially overcome in experiments measuring the joint spectral~\cite{kim2005measurement,wasilewski2006joint,avenhaus2009fiber,zielnicki2018joint} or temporal~\cite{kuzucu2008joint,odonnell2009time,maclean2018direct} intensity of the photon pairs.
However, these measurements are insensitive to correlations in the spectral phase of the photons.
Phase correlations arise when spectrally-structured pump fields are used to generate time-frequency entangled photons~\cite{ansari2018tomography} which have applications in quantum communication~\cite{tittel2000quantum, nunn2013large} and sensing~\cite{lamine2008quantum,dorfman2016nonlinear,lyons2018attosecond,ansari2021achieving}.
They also arise when the pump is chirped and in such cases can degrade the purity of heralded single photons as well as the interference quality between independent sources~\cite{u2006generation,bell2020diagnosing}.
Such phase correlations must be minimized when preparing the high-purity photons needed for photonic networks~\cite{meyer2017limits} and information processors~\cite{zhong2020quantum,arrazola2021quantum}.
Finally, the joint spectral phase's structure can reveal interesting physics in the photon pair creation process in both atomic systems~\cite{du2008narrowband} and nonlinear crystals~\cite{triginer2020understanding}.
All these applications benefit from a full characterization of the time-frequency structure of photon pairs.
Recent experiments have demonstrated this either by employing nonlinearities~\cite{maclean2019reconstructing,davis2020measuring} or linear techniques that are only applicable to highly correlated pair sources~\cite{beduini2014interferometric,chen2015measuring,tischler2015measurement,jizan2016phase}.

In this paper, we propose a general scheme to determine the spectral mode of light and demonstrate it experimentally using photon pairs.
We interfere the photons with a weak reference pulse and measure spectral intensity correlations.
Our scheme does not require nonlinearities, phase stability, spectral shaping, or complex computational algorithms.
It does however require \textit{a priori} knowledge of the spectral mode of the reference pulse.
Because the reference is simply an attenuated laser, its mode can be measured using conventional self-referencing techniques for classical pulses~\cite{kane1993characterization,iaconis1998spectral}.

\begin{figure}
    \centering
    \includegraphics[width=1\columnwidth]{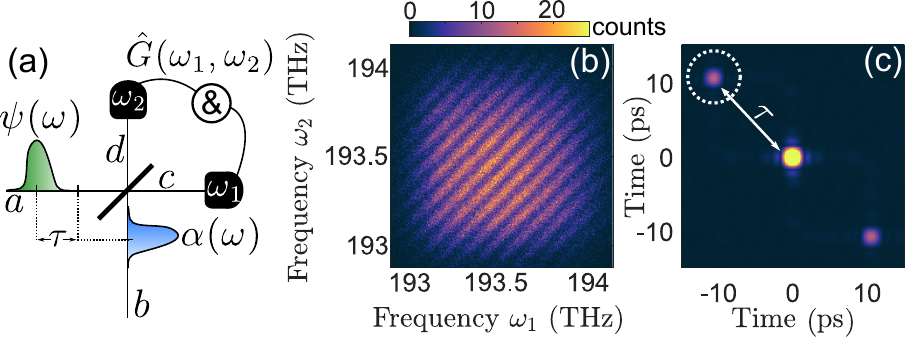}
	\caption{(a) We combine the unknown signal $\psi(\omega)$ with a reference $\alpha(\omega)$ on a BS and measure spectral intensity correlations $\hat{G}(\omega_1,\omega_2)$ [Eq.~\eqref{eqn:cross_correlation}].
	(b) Example of an experimentally measured $\braket{\hat{G}(\omega_1,\omega_2)}$.
	(c) Fourier transform of $\braket{\hat{G}(\omega_1,\omega_2)}$. 
	The interferometric term $\Gamma(\omega_1, \omega_2)$ [Eq.~\ref{eqn:int_term}] is isolated using a window function. 
	White dotted line shows 50\% contour of a Gaussian window function.
	}
	\label{fig:scheme}
\end{figure}
 
We begin by describing the single photon case which is extended later to the bi-photon case.
We consider for now a signal in a pure single photon pulse $\int d\omega \psi(\omega) \hat{a}^\dagger_\omega \ket{0}$ where $\hat{a}^\dagger_\omega$ is a creation operator at frequency $\omega$ in input mode $a$.
The goal is to determine its spectral mode $\psi(\omega)$.
As shown in Fig.~\ref{fig:scheme}(a), the photon is combined with a reference pulse on a beam splitter (BS).
The input state of the BS is:
\begin{equation}
\ket{\Psi} = \int d\omega \psi(\omega) \hat{a}^\dagger_\omega \ket{0} \otimes \int d\omega' \ket{\alpha(\omega')},
\label{eqn:two_photons}
\end{equation}
where we assumed the reference is a coherent state with amplitude $\alpha(\omega)$, i.e. $\hat{b}_{\omega_0} \int d\omega' \ket{\alpha(\omega')} = \alpha(\omega_0) \int d\omega' \ket{\alpha(\omega')}$.
At the output of the BS, one measures spectral intensity correlations which are described by the observable
\begin{equation}
    \hat{G}(\omega_1,\omega_2) \equiv \hat{d}^\dagger_{\omega_2} \hat{c}^\dagger_{\omega_1} \hat{c}_{\omega_1} \hat{d}_{\omega_2}.
    \label{eqn:cross_correlation}
\end{equation}
By repeating measurements of Eq.~\eqref{eqn:cross_correlation}, one determines the expectation value $\braket{\hat{G}(\omega_1,\omega_2)}$ which is the second-order cross-correlation function of the two output fields evaluated at $\omega_1$ and $\omega_2$.
We can compute this expectation value with respect to the input state in Eq.~\eqref{eqn:two_photons} by using the BS input-output transformations, $\hat{c}_\omega = (\hat{a}_\omega +\hat{b}_\omega)/\sqrt{2}$ and $\hat{d}_\omega = (\hat{a}_\omega -\hat{b}_\omega)/\sqrt{2}$:
\begin{equation}
\begin{split}
    \braket{\hat{G}(\omega_1,\omega_2)} &= \braket{\Psi|\hat{G}(\omega_1,\omega_2)|\Psi} \\
    &= \frac{1}{4}\left| \alpha(\omega_1)\psi(\omega_2) - \psi(\omega_1)\alpha(\omega_2)\right|^2 \\
    &\qquad + \frac{1}{4}|\alpha(\omega_1)\alpha(\omega_2)|^2.
\end{split}
    \label{eqn:hom}
\end{equation}
When the reference intensity is at the single-photon level, a measurement of Eq.~\eqref{eqn:hom} reveals an interference pattern which depends on both the amplitude and phase of the spectral mode of the input fields [Fig.~\ref{fig:scheme}(b)].
Similar interference patterns have been measured in experiments performing spectrally-resolved Hong-Ou-Mandel interferometry, where both inputs are single photons~\cite{jin2015spectrally, gerrits2015spectral, thiel2020single,zhang2021high}.
%In our case, the interference visibility is reduced by the two-photon component of the reference, i.e. $|\alpha(\omega_1)\alpha(\omega_2)|^2$~\cite{rarity2005non}.

In order to isolate the mode function of the signal from Eq.~\eqref{eqn:hom}, we use a Fourier filtering technique~\cite{froehly1973time, fittinghoff1996measurement,beck2001joint,thiel2020single}.
That is, we delay the reference by $\tau$, i.e. $\alpha(\omega)\rightarrow \alpha(\omega)e^{i\omega\tau}$. 
Expanding Eq.~\eqref{eqn:hom}, we obtain:
\begin{equation}
\begin{split}
     \braket{\hat{G}(\omega_1,\omega_2)}
    &= \frac{1}{4}\left(\zeta(\omega_1, \omega_2) - \Gamma(\omega_1, \omega_2) -\Gamma^*(\omega_1, \omega_2) \right).
\end{split}
\label{eqn:photonsG}
\end{equation}
The first term, $\zeta(\omega_1, \omega_2) = \left|\alpha(\omega_1)\psi(\omega_2) \right|^2 + \left|\psi(\omega_1)\alpha(\omega_2) \right|^2 + |\alpha(\omega_1)\alpha(\omega_2)|^2$, depends on the spectral amplitudes of the fields but not their spectral phases. 
The second term,
\begin{equation}
\Gamma(\omega_1, \omega_2) =  \psi(\omega_1)\psi^*(\omega_2)\alpha^*(\omega_1)\alpha(\omega_2)e^{i(\omega_2-\omega_1)\tau},
\label{eqn:int_term}
\end{equation}
depends on both quantities.
This interference term can be isolated in the Fourier domain.
Namely, by performing a two-dimensional Fourier transform $\mathcal{F}$ of Eq.~\eqref{eqn:photonsG}, one finds that $\mathcal{F}\left\{\Gamma(\omega_1, \omega_2)\right\}$ and $\mathcal{F}\left\{\Gamma^*(\omega_1, \omega_2)\right\}$ are symmetrically separated from $\mathcal{F}\left\{\zeta(\omega_1, \omega_2)\right\}$ by the temporal delay $\tau$ [Fig.~\ref{fig:scheme}(c)].
If $\tau$ is made larger than the temporal duration of the signal and reference pulses, one can isolate $\Gamma(\omega_1, \omega_2)$ by multiplying $\mathcal{F}\left\{\braket{\hat{G}(\omega_1,\omega_2)}\right\}$ by a window function (e.g. Gaussian) centered on $\mathcal{F}\left\{\Gamma(\omega_1, \omega_2)\right\}$ and taking the inverse Fourier transform~\cite{SM}.
One can then divide the filtered interferogram by the reference spectral mode and a phase term depending on the delay $\tau$ which are both \textit{a priori} known quantities, i.e. $\psi(\omega_1)\psi^*(\omega_2) = \Gamma(\omega_1, \omega_2) / \alpha^*(\omega_1)\alpha(\omega_2)e^{i(\omega_2-\omega_1)\tau}$.
This division step imposes that the reference pulse should be at least as spectrally broad as the signal, and thus a characterization of the signal and reference spectra prior to the measurement is required to verify that this condition is met.
There are otherwise no special requirements for the reference.
The pure signal spectral mode can then be obtained by diagonalizing the matrix $\Phi(\omega_1, \omega_2) = \psi(\omega_1)\psi^*(\omega_2)$, where $\omega_1$ and $\omega_2$ are discrete measurement bins.
Suppose instead the signal is in a mixed state of modes $\Phi(\omega_1,\omega_2) = \sum_i p_i\Phi_i(\omega_1,\omega_2)$.
Then one can show that the Fourier filtered $\braket{\hat{G}(\omega_1,\omega_2)}$ is determined by $\sum_i p_i \Gamma_i(\omega_1, \omega_2)$ where $\Gamma_i(\omega_1, \omega_2)$ is given by Eq.~\eqref{eqn:int_term} with $\psi(\omega_1)\psi^*(\omega_2) \rightarrow \Phi_i(\omega_1,\omega_2)$.
Then, $\Phi(\omega_1,\omega_2) = \sum_i p_i \Gamma_i(\omega_1, \omega_2) / \alpha^*(\omega_1)\alpha(\omega_2)e^{i(\omega_2-\omega_1)\tau}$ and thus modal mixtures can also be obtained without further measurements.
%Such a measurement was experimentally demonstrated in Ref.~\cite{thiel2020single} where one of the two photons served as the reference field.

%Thus far, we assumed that both the signal and reference are single photons.
%But the scheme applies to any state of light, including classical ones.
The method outlined above is not restricted to single photons and can be used to determine the spectral mode of any quantum or classical state of light.
%By forgoing the final diagonalization step, it generalizes to measuring statistical mixtures of spectral modes like $\rho(\omega_1,\omega_2) = \sum_i p_i \psi_i(\omega_1)\psi_i(\omega_2)$~\cite{thiel2020single}.
Moreover, since it measures spectral intensity correlations rather than the spectral intensity alone, there does not need to be any phase coherence between the reference and signal~\cite{dorrer2003linear}, e.g. these can be phase-randomized coherent states or independent thermal states as in Hanbury Brown and Twiss interferometry.
%Since it relies on measuring spectral intensity correlations rather than the spectral intensity alone, the signal and references do not need any first-order coherence, e.g. 
Let us consider the former case as an example.
One can replace the creation and anihilation operators in Eq.~\eqref{eqn:cross_correlation} with the corresponding spectral amplitudes, i.e. $\hat{c}_\omega \rightarrow \left(\psi(\omega)+\alpha(\omega)e^{i\omega \tau}\right)/\sqrt{2}$ and $\hat{d}_\omega \rightarrow \left(\psi(\omega)-\alpha(\omega)e^{i\omega \tau}\right)/\sqrt{2}$.
Inserting these transformations into Eq.~\eqref{eqn:cross_correlation} and taking the classical ensemble average, denoted by $\braket{}_\mathcal{C}$, one finds:
\begin{equation}
\begin{split}
    \braket{\hat{G}(\omega_1,\omega_2)}_\mathcal{C} = \frac{1}{4} &[\zeta(\omega_1, \omega_2) + \left|\psi(\omega_1)\psi(\omega_2) \right|^2  \\
    &\quad - \Gamma(\omega_1, \omega_2) -\Gamma^*(\omega_1, \omega_2)  ].
\end{split}
\label{eqn:coherentG}
\end{equation}
Comparing with Eq.~\eqref{eqn:photonsG} where we assumed the signal to be a single photon, an additional spectral-phase-insensitive term $\left|\psi(\omega_1)\psi(\omega_2) \right|^2$ appears in Eq.~\eqref{eqn:coherentG} due to the intensity fluctuations of the signal, now assumed to be a coherent state.
This additional term further limits the visibility of the interference fringes in $\braket{\hat{G}(\omega_1,\omega_2)}_\mathcal{C}$.
In principle, the visibility is limited to 50\% for phase-randomized coherent states~\cite{scully1999quantum}.
A reduced visibility is not an issue so long as $\Gamma(\omega_1, \omega_2)$ is distinguishable from the spectral-phase-insensitive terms and any other noise in the Fourier domain, which might require averaging $ \braket{\hat{G}(\omega_1,\omega_2)}$ over a longer period of time. 

We now extend the scheme to measure the bi-photons produced by processes like spontaneous parametric down-conversion (SPDC) or four-wave mixing:
\begin{equation}
    \ket{\mathrm{SPDC}} = \iint d\omega_1 d\omega_2 f(\omega_1, \omega_2) \hat{a}^\dagger_{\omega_1} \hat{h}^\dagger_{\omega_2}\ket{0}.
    \label{eqn:spdc}
\end{equation}
Here, $f(\omega_1, \omega_2)$ is termed the joint spectral amplitude (JSA) and characterizes the joint spectral mode of the photon pair.
Suppose one performs spectrally-resolved detection in the herald mode $h$.
By detecting a photon of frequency $\omega_h$, one heralds a signal photon in mode $a$ whose spectral mode is given by $\psi(\omega_1) = f(\omega_1, \omega_h)$.
The heralded photon can then be combined with a reference in order to measure $f(\omega_1, \omega_h)$ using the aforementioned procedure.
The quantity $f(\omega_1, \omega_h)$ is a cross-section of the JSA along $\omega_2 = \omega_h$ and hence this measurement should be repeated for all herald frequencies in order to determine $f(\omega_1, \omega_2)$.
One caveat is that there is a phase between each cross-section which remains undetermined because of the spectral-phase-insensitive detection in mode $h$.
%Since this phase is uncorrelated to the signal, it does not affect the modal purity of the SPDC source.
However, one can repeat the measurement using mode $a$ as the herald and mode $h$ as the signal to be combined with the reference.
These two measurements unambiguously determine the full JSA~\cite{davis2020measuring}.

%\section{Experiment}

\begin{figure}[t]
    \centering
    \includegraphics[width=1\columnwidth]{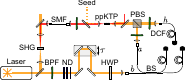}
	\caption{Experimental setup. The SMF and seed are included only in certain measurements which are described in the main text. SHG: second-harmonic generation, SMF: single-mode fiber, BPF: bandpass filter, ND: neutral-density filter, ppKTP: periodically poled potassium titanyl phosphate, (P)BS: (polarizing) beam splitter, HWP: half-wave plate, DCF: dispersion-compensating fiber.
	}
	\label{fig:setup}

\end{figure}

We now turn to our experiment.
The experimental setup is shown in Fig.~\ref{fig:setup}.
An optical parametric oscillator produces pulses (150 fs duration, 1550 nm center wavelength) at a repetition rate of 80 MHz.
A small fraction of the power is used as the reference while the remaining power is used to prepare the pump light for the SPDC source.
%We assume the reference pulses are chirpless and use a spectrometer to measure the real-valued $\alpha(\omega)$.
The pump pulses are frequency-doubled in a lithium niobate crystal and subsequently coupled into a 8-mm-long periodically-poled potassium titanyl phosphate (ppKTP) waveguide.
A type-II SPDC interaction inside the waveguide produces pairs of photons described by Eq.~\eqref{eqn:spdc}.
%Bandpass filters are used to eliminate the sinc-sidelobes of the down-converted photons but not filter its main feature.
Our goal is to measure both the amplitude and phase of the JSA $f(\omega_1, \omega_2)$.

We send the down-converted photon in mode $h$ directly into a spectrally-resolving single photon detector.
The heralded photon in mode $a$ is combined with the reference pulse in a single-mode fiber BS.
We use a motorized stage to set a delay of $\tau = 10.00(7)$ ps between the two pulses.
We then perform spectrally-resolved detection at the output of the BS.
Each spectrally-resolving single photon detector consists of a dispersion-compensating fiber (DCF) having a dispersion of -997 ps/nm followed by a superconducting nanowire detector.
The DCF maps the photon's frequency to its arrival time at the detector which is recorded using a time-tagging device.
The combined detector and time-tagging temporal jitter is roughly 40 ps resulting in a spectral uncertainty of 40 pm (5 GHz).
The combined transmission and detection efficiency of each path is approximately 3\% and is mainly limited by the transmission of the DCFs (15\%).

We measure roughly $10^5$ single photons per second from the SPDC source using 3 mW of pump power.
With the reference having approximately $10^6$ photons per second, we measure three-fold coincidence events at a rate of about $100$ per second.
We acquire data for a few hours and obtain a three-dimensional histogram $N(\omega_1,\omega_2,\omega_h)$ which is determined by the joint probability to measure frequencies $(\omega_1, \omega_2)$ at the output of the BS and $\omega_h$ in the herald mode.
In order to determine $f(\omega_1, \omega_2)$, we process $N(\omega_1,\omega_2,\omega_h)$ in the following manner.
Firstly, the measured frequencies are placed into discrete bins each having a width of approximately $10$ GHz.
For the $j$th herald bin, $N(\omega_1,\omega_2,\omega_{h_j}) = \braket{\hat{G}_{j}(\omega_1,\omega_2)}$ is the cross-correlation function conditioned on having detected the herald photon in the frequency bin $\omega_{h_j}$.
An example of this quantity is shown in Fig.~\ref{fig:scheme}(b).
Secondly, we use $\braket{\hat{G}_{j}(\omega_1,\omega_2)}$ to determine $f(\omega_1, \omega_{h_j})$ using the Fourier filtering procedure.
This process is repeated for all the herald bins and we obtain the full $f(\omega_1, \omega_2)$.
The reference pulse spectrum $|\alpha(\omega)|^2$ is measured by blocking the down-converted photons.
We assume that the reference photons are approximately chirpless and so take $\alpha(\omega)$ to be real-valued.
%could justify this assumption with the fact that our beta measurement agrees with the SPIDER measurement which is self-referencing.

\begin{figure}
    \centering
    \includegraphics[width=1\columnwidth]{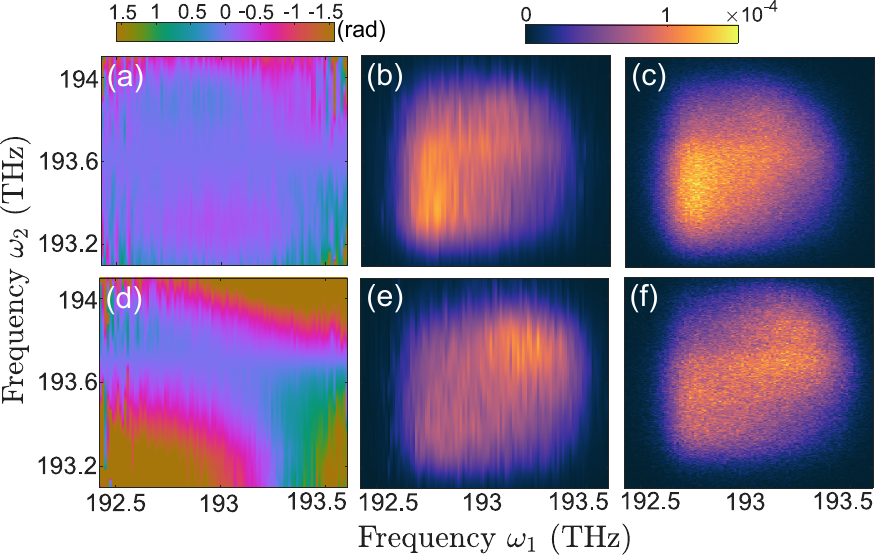}
	\caption{Measured joint spectral amplitude. Top (bottom) row shows results with a chirpless (chirped) pump. (a) and (d) are the joint spectral phases $\arg \{ f(\omega_1,\omega_2) \}$ while (b) and (e) are the amplitudes $|f(\omega_1,\omega_2)|$.
	(c) and (f) are the amplitudes measured using a conventional phase-insensitive method.
	}
	\label{fig:results}
\end{figure}

As a first test, we adjust the pump bandwidth using a bandpass filter so that the down-converted photons have an uncorrelated JSA, i.e. $f(\omega_1,\omega_2)= f_1(\omega_1)f_2(\omega_2)$.
Bandpass filters are also used after the waveguide to eliminate the sinc-sidelobes from the down-converted spectra.
We plot the measured $\arg \{ f(\omega_1,\omega_2) \}$ and $|f(\omega_1,\omega_2)|$ in Fig.~\ref{fig:results}(a) and (b), respectively.
The latter can be compared with a conventional measurement of $|f(\omega_1,\omega_2)|$ where we block the reference and record two-fold coincidences [Fig.~\ref{fig:results}(c)].
To quantify the degree of correlation between the down-converted photons, we perform a Schmidt decomposition of the complex JSA and obtain a Schmidt number of $K=1.02$.
This number is close to unity which indicates that the down-converted photons are indeed uncorrelated in time-frequency.
We also measure the second-order autocorrelation function $g^{(2)}$ of the signal and herald modes with the reference blocked. 
We find $1.84(2)$ and $1.85(2)$, respectively.
For an idealized photon pair source, $g^{(2)}_{\mathrm{ideal}} = 1+1/K$~\cite{christ2011probing}.
However, in addition to photon pairs, our source also generates uncorrelated single photons (4(2)\% of the total counts) due to unguided down-conversion processes in the waveguide.
Subtracting the noise photons from the $g^{(2)}$ calculation~\cite{SM,eckstein2011realistic}, we find $1.98(7)$ and $1.99(7)$ in the signal and herald modes, respectively, which agree with the value expected from the Schmidt number, $g^{(2)}_{\mathrm{ideal}} \approx 1.98$.

For a second test, we chirp the pump pulse by coupling it into 5m-long single mode fiber.
We characterize the chirp using spectral phase interferometry for direct electric-field reconstruction (SPIDER)~\cite{iaconis1998spectral}.
The chirped pulse is well-described by a quadratic spectral phase, i.e $|A(\omega_p)|e^{-i \frac{\beta}{2} \omega^2_p}$ where $|A(\omega_p)|$ is the spectral amplitude and $\beta = 2.0(4) \times 10^{5}~\mathrm{fs}^2$ is the measured group delay dispersion parameter.
The pump chirp introduces a correlated phase in the down-converted photons, $f(\omega_1,\omega_2)= f_1(\omega_1)f_2(\omega_2)e^{-i\beta\omega_1\omega_2}$.
This correlated phase is visible in the measured $\arg \{ f(\omega_1,\omega_2) \}$ [Fig.~\ref{fig:results}(d)].
Fitting a quadratic function, we find $\beta = 1.69(2) \times 10^{5}~\mathrm{fs}^2$ which agrees with the aforementioned value.
The slight difference in $|f(\omega_1,\omega_2)|$ [Fig.~\ref{fig:results}(e)] compared to the chirpless pump case [Fig.~\ref{fig:results}(b)] is due to self-phase modulation which modifies the pump's spectral amplitude as it propagates inside the fiber.
This difference is also apparent in the conventional phase-insensitive measurements [Fig.~\ref{fig:results}(c),(f)].
The Schmidt number of the measured complex JSA is $K=1.48$ while it is $K=1.04$ if one ignores the phase.
Thus, the time-frequency correlations of the down-converted photons are mainly caused by the non-uniform spectral phase of the pump.
Due to these correlations, the noise-subtracted $g^{(2)}$ of the signal and herald modes decreases to 1.62(6) and 1.68(6), respectively, which agree with the expected $g^{(2)}_{\mathrm{ideal}} \approx 1.67$.

\begin{figure}[t]
    \centering
    \includegraphics[width=1\columnwidth]{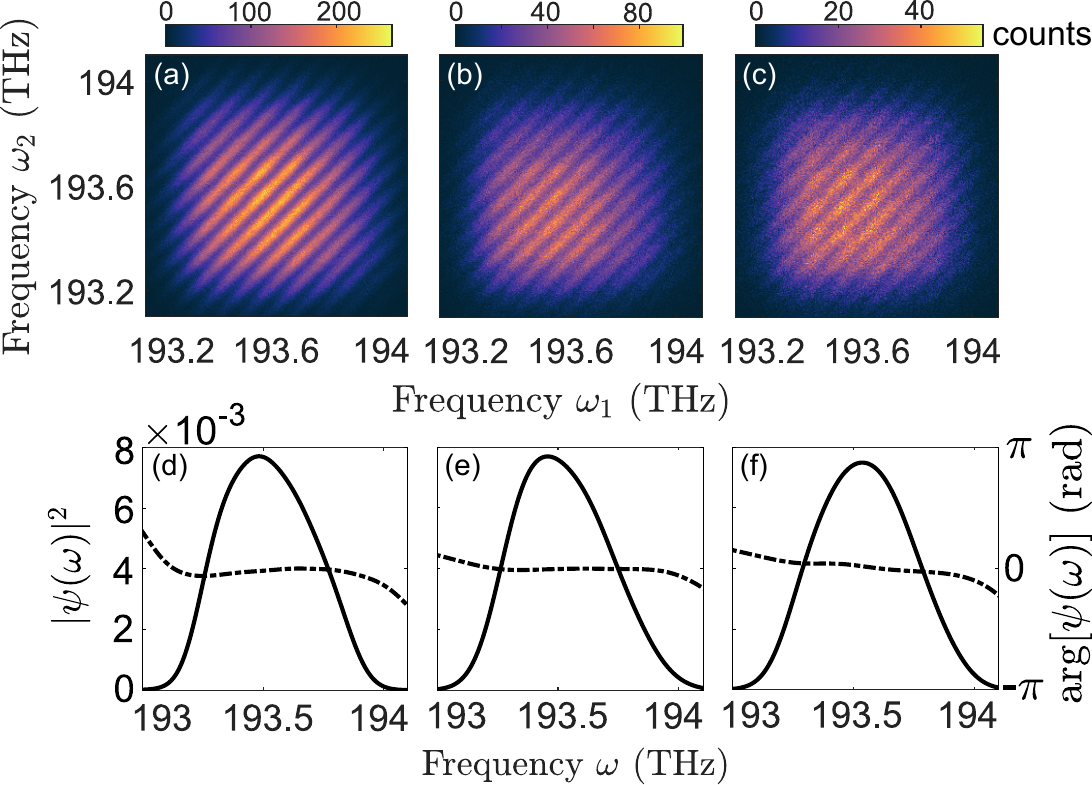}
	\caption{$\braket{\hat{G}(\omega_1,\omega_2)}$ measured using a signal with (a) sub-Poissonian, $g^{(2)} = 0.57(4)$ (b) Poissonian, $g^{(2)} = 1.048(3)$, (c) super-Poissonian, $g^{(2)} = 1.84(2)$ photon statistics. 
	In all three cases, the reference is a coherent state with Poissonian statistics.
	The fringe visibilities are $0.27(2)$, $0.17(1)$, $0.14(1)$, respectively.
	Bottom row of plots shows the corresponding $|\psi(\omega)|^2$ (left $y$-axis scale, solid line) and $\arg[\psi(\omega)]$ (right $y$-axis scale, dashed line).
	}
	\label{fig:example_specs}
\end{figure}

So far, we demonstrated that our scheme can be used to measure both amplitude and phase of the JSA of photon pairs produced by SPDC.
One potential drawback of our measurement is that it relies on measuring three-fold coincidences which can lead to slow data acquisition with pair sources that are faint or have low heralding efficiencies. % especially given the typically poor efficiency of spectrally-resolved single photon detection. 
%Talk about heralding efficiency thing here?
To resolve this issue, we propose a technique to measure the JSA that uses bright classical fields and hence can be much quicker.
The technique draws inspiration from stimulated emission tomography~\cite{liscidini2013stimulated}.
One couples a continuous-wave seed laser with tunable frequency $\omega_s$ into the pair source.
Through difference frequency generation with the pump, the signal is prepared in a bright coherent state.
The spectral mode of this stimulated signal is given by the cross-section of the JSA at the seed frequency $\omega_s$, i.e. $f(\omega, \omega_s)$.
One can then combine the stimulated signal with the reference and measure $\braket{\hat{G}(\omega_1,\omega_2)}$ to determine $f(\omega, \omega_s)$, and repeat this process for different seed frequencies $\omega_s$.
Since the signal and reference do not need to be phase-stable, it is not necessary to lock the seed and pump lasers.

We perform a proof-of-principle demonstration of this technique.
In addition to the unchirped pump, we couple a seed beam into the SPDC source whose polarization is aligned with the herald mode. 
The seed beam is produced by an attenuated continuous-wave laser (1560 nm, 192 THz).
%The signal-to-noise of the stimulated signal is about $10^3$, which is obtained by dividing the number of photons in the signal mode with both the pump and seed unblocked by the number of photons when either of these is blocked.
We combine the stimulated signal with the reference pulse on a BS and measure $\braket{\hat{G}(\omega_1,\omega_2)}$ by recording two-fold coincidences.
The result is shown in Fig.~\ref{fig:example_specs}(b).
The fringe visibility is reduced compared to the heralded measurement [Fig.~\ref{fig:example_specs}(a)] due to increased intensity fluctuations of the signal.
The benefit is that $\braket{\hat{G}(\omega_1,\omega_2)}$ can be measured much more quickly.
Using approximately $10^6$ photons per second in both the reference and signal (i.e. 0.01 photons per pulse), we obtain $10^4$ two-fold coincidences per second.
This rate was limited by the dead time and dynamic range of the single photon detectors.
One could in principle measure $\braket{\hat{G}(\omega_1,\omega_2)}$ even more quickly by measuring shot-by-shot correlations between two spectrometers employing regular photodetectors.

For the sake of demonstration, we also measure $\braket{\hat{G}(\omega_1,\omega_2)}$ when there is thermal noise in the signal by turning off the seed beam and recording two-fold coincidences.
This last measurement ignores the herald photon and hence the signal has super-Poissonian photon statistics which further reduces the fringe visibility [Fig.~\ref{fig:example_specs}(c)].
We determine the spectral modes in the heralded, seeded, and unseeded cases using the Fourier filtering procedure [Fig.~\ref{fig:example_specs}(d)-(f)].
The three spectral modes have an average pairwise fidelity $\left|\int d\omega \psi_1(\omega)\psi_2^*(\omega)\right|$ of 0.991(4) which demonstrates that our measurement is insensitive to the fringe visibility due to the Fourier filtering.

In summary, we demonstrated a scheme that can determine the joint spectral mode of the photons pairs produced by SPDC or four-wave mixing.
By using a combination of intensity interferometry and Fourier filtering, our scheme is resilient to phase instabilities and intensity fluctuations.
An analogous scheme measuring spatial intensity correlations can be used to characterize light's spatial mode~\cite{chrapkiewicz2016hologram,defienne2021polarization}.
Finally, extending the scheme beyond two photons should be possible by combining each photon with a reference and measuring spectral intensity correlations across all modes.  

\begin{acknowledgements}
We thank Single Quantum for loaning us superconducting nanowire detectors.
This work was supported by: Engineering and Physical Sciences Research Council (P510257); H2020 Marie Sklodowska-Curie Actions (846073); Korea Institute of Science and Technology open research program; National Research Council of Canada.
\end{acknowledgements}

\bibliographystyle{apsrev4-2}
\bibliography{refs}

%%% SM %%%

% UNCOMMENT FOR ARXIV

\newpage
\onecolumngrid
\input{./sm/sm.tex}

\end{document}

%% file: sm/sm.tex
%% UNCOMMENT FOR SEPARATE DOCUMENT

% \documentclass[amsmath,amssymb,reprint,aps,prl,floatfix,superscriptaddress]{revtex4-2}
% \usepackage{bm}
% \usepackage{amsmath}
% \usepackage{graphicx}
% \usepackage{braket}
% \usepackage{bbm}
% \usepackage[hidelinks]{hyperref}

% \begin{document}

% \title{Supplemental material for ``Measuring the joint spectral mode of photon pairs using intensity interferometry"}

% \author{G.S. Thekkadath}
% \email{g.thekkadath@imperial.ac.uk}
% \affiliation{Department of Physics, Imperial College London, Prince Consort Rd, London SW7 2AZ, UK}
% \affiliation{National Research Council of Canada, 100 Sussex Drive, Ottawa, K1A 0R6, Canada}

% \author{B.A. Bell}
% \affiliation{Department of Physics, Imperial College London, Prince Consort Rd, London SW7 2AZ, UK}

% \author{R.B. Patel}
% \affiliation{Department of Physics, Imperial College London, Prince Consort Rd, London SW7 2AZ, UK}
% \affiliation{Clarendon Laboratory, University of Oxford, Parks Road, Oxford, OX1 3PU, UK}

% \author{M.S. Kim}
% \affiliation{Department of Physics, Imperial College London, Prince Consort Rd, London SW7 2AZ, UK}

% \author{I.A. Walmsley}
% \affiliation{Department of Physics, Imperial College London, Prince Consort Rd, London SW7 2AZ, UK}

% \maketitle

% \onecolumngrid

%% UNCOMMENT FOR ARXIV

\renewcommand{\theequation}{S\arabic{equation}}
\renewcommand{\thefigure}{S\arabic{figure}}
\section*{Supplemental material}

\subsection*{Window function}
We employ a Fourier filtering technique to isolate the interference term $\Gamma(\omega_1, \omega_2)$ from the measured $\braket{\hat{G}(\omega_1,\omega_2)}$.
This is achieved by multiplying $\mathcal{F}\{\braket{\hat{G}(\omega_1,\omega_2)}\}$ by a window function $W(t_1, t_2)$ centered on $\mathcal{F}\{ \Gamma(\omega_1, \omega_2) \}$, as shown in Fig.~1(c) of the main text.
%such as a Gaussian: $W(t_1, t_2) = \exp{\left( \frac{-(t_1-t_{01})^2 - (t_2-t_{02})^2 }{2\sigma^2}\right)}$ where $t_{01}$ and $t_{02}$ are chosen to center the Gaussian on $\mathcal{F}\{ \Gamma(\omega_1, \omega_2) \}$, as shown in Fig.~1(c)
%This is achieved by multiplying $\mathcal{F}\{\braket{\hat{G}(\omega_1,\omega_2)}\}$ by a Gaussian window-function $W(t_1, t_2) = \exp{\left( \frac{-(t_1-t_{01})^2 - (t_2-t_{02})^2 }{2\sigma^2}\right)}$ where $t_{01}$ and $t_{02}$ are chosen to center the Gaussian on $\mathcal{F}\{ \Gamma(\omega_1, \omega_2) \}$, as shown in Fig.~1(c).
Taking the inverse Fourier transform $\mathcal{F}^{-1}$ of the result, we obtain:
\begin{equation}
    \mathcal{F}^{-1}\left\{ W(t_1, t_2) \mathcal{F}\{\braket{\hat{G}(\omega_1,\omega_2)}\} \right\} = \tilde{W} * \braket{\hat{G}(\omega_1,\omega_2)} \approx \tilde{W} * \Gamma(\omega_1,\omega_2) \approx \Gamma(\omega_1,\omega_2)
    \label{eqn:convolution}
\end{equation}
where $*$ denotes a two-dimensional convolution and $\tilde{W}$ is the inverse Fourier transform of $W$.
The first approximation in Eq.~\eqref{eqn:convolution} is valid when the window-function is approximately zero in the region around $\mathcal{F}\left\{\zeta(\omega_1, \omega_2)\right\}$ and  $\mathcal{F}\left\{\Gamma^*(\omega_1, \omega_2)\right\}$.
The second approximation is valid when the window function is approximately unity in the region around $\mathcal{F}\{ \Gamma(\omega_1, \omega_2) \}$.

\begin{figure}
    \centering
    \includegraphics[width=0.85\columnwidth]{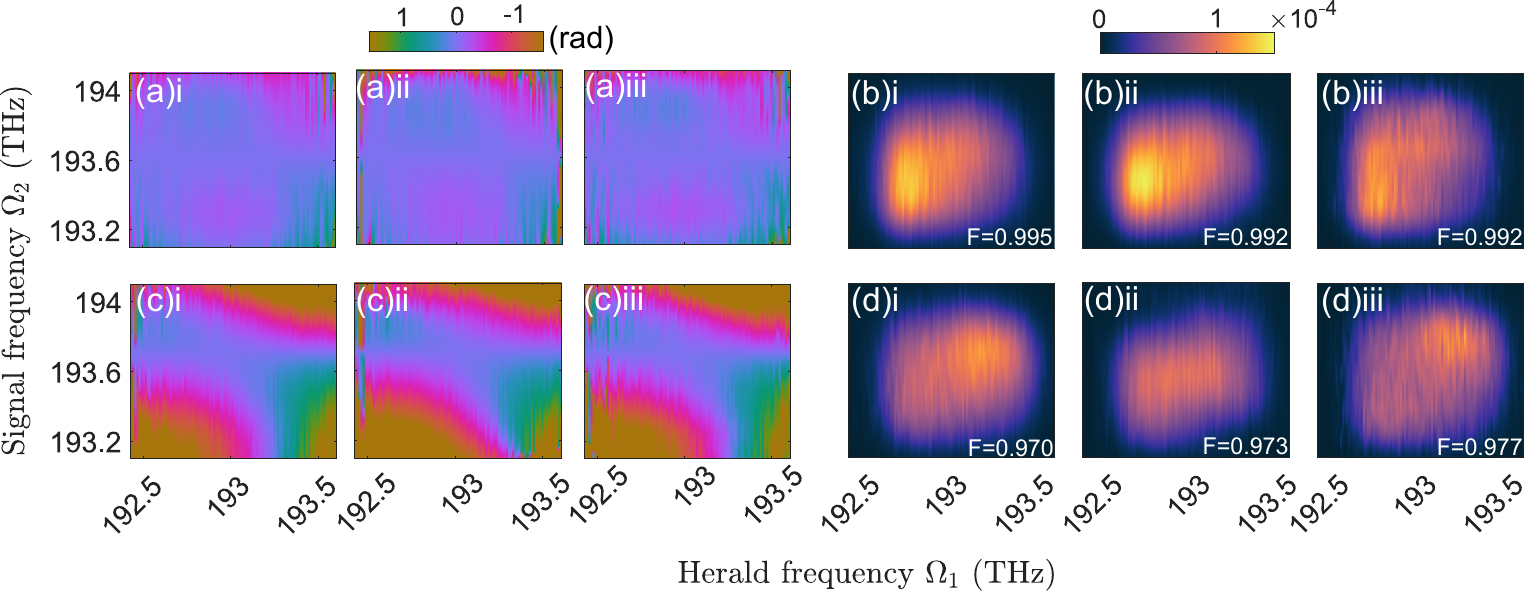}
	\caption{Joint spectral amplitudes obtained using three different window functions. Case (i) and (iii) uses a Gaussian with standard deviation $\sigma=1.7$~ps and $\sigma=2.5$~ps, respectively.
	Case (ii) uses a rectangular window of width $\sigma=2.5$~ps.
	(a) and (c) are the joint spectral phases $\arg \{ f(\omega_1,\omega_2) \}$ while (b) and (e) are the amplitudes $|f(\omega_1,\omega_2)|$ in the unchirped and chirped cases, respectively. 
	}
	\label{fig:comparingWindows}
\end{figure}

In Fig.~\ref{fig:comparingWindows}, we show the joint spectral amplitudes obtained using various window functions.
In cases (i) and (iii), we employ a Gaussian window  $W(t_1, t_2) = \exp{\left( \frac{-(t_1-t_{01})^2 - (t_2-t_{02})^2 }{2\sigma^2}\right)}$ with standard deviation $\sigma=1.7$~ps and $\sigma=2.5$~ps, respectively.
In cases (ii), we employ a rectangular window $W(t_1, t_2) = \mathrm{rect}{\left( \frac{t_1-t_{01}}{\sigma}\right)}\times\mathrm{rect}{\left( \frac{t_2-t_{02}}{\sigma}\right)}$ of width $\sigma=2.5$ ps.
%The window size does not significantly change the overall features of the joint spectral amplitude or phase.
The results obtained with the smaller window size have fewer artifacts due to stronger low-pass filtering.%, but the overall features of all three distributions are similar. 

We also quantify the overlap between the joint spectral intensities obtained using our Fourier filtering procedure with the ones obtained directly via the conventional phase-insensitive method (i.e. without interfering with the reference).
The overlap between these two distributions, $F(p,q) = \sum_i\sqrt{p_iq_i}$, is shown in the bottom right of Fig.~\ref{fig:comparingWindows}(b),(d).
The large overlap shows that the window-function convolution in Eq.~\eqref{eqn:convolution} has a negligible effect in all three cases.
The data presented in the main text uses the window function of case (iii).
% We can verify the validity of these approximations by comparing the joint spectral amplitude obtained using our Fourier filtering procedure with the one obtained directly via the conventional phase-insensitive method (i.e. without interfering with the reference).
% We calculate the overlap between the distributions using the total variational distance, $F(p,q) = \sum_i\sqrt{p_iq_i}$.  
% The overlap between Fig.~3(b) and Fig.~3(c) in the main text is 0.98 (unchirped case), while between Fig.~3(e) and Fig.~3(f), it is 0.95 (chirped case).
% The high overlap between these distributions suggests that the convolution from the window-function is not a significant effect.

% We also show the joint spectral amplitude and phase obtained using two different window sizes $\sigma$ in Fig.~\ref{fig:comparingWindows}.
% In cases (ii), the window size is 50\% larger than in cases (i).
% The window size does not significantly change the overall features of the joint spectral amplitude or phase.
% However, the distributions obtained with a smaller window size have somewhat fewer artifacts due to stronger low-pass filtering.

\subsection*{Noise photons}

In addition to photon pairs, our source produces uncorrelated noise photons over a broad spectrum due to unguided down-conversion processes in the waveguide.
These noise photons are filtered out by the coincidence measurement used to obtain the joint spectral amplitude.
However, they appear when measuring photons from only one of down-converted modes such as when we measure the second-order autocorrelation function $g^{(2)}$ of the signal and herald modes.
As a result, the $g^{(2)}$ values we measure are lower than the values expected from the Schmidt number $K$ of the joint spectral amplitude, $g^{(2)}_{\mathrm{ideal}} = 1+1/K$.

\begin{figure}
    \centering
    \includegraphics[width=0.5\columnwidth]{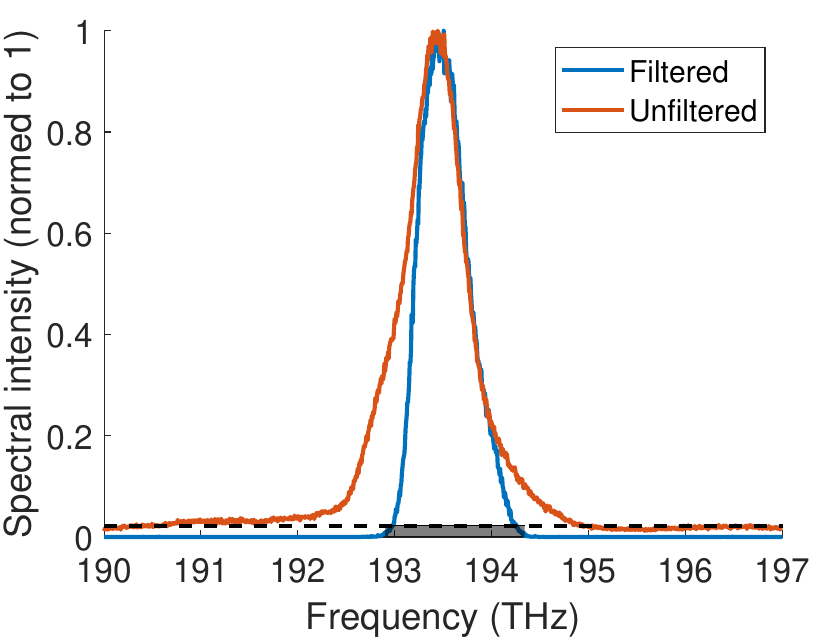}
	\caption{Marginal spectrum of signal photons measured with [blue] and without [orange] bandpass filter.
	The flat background in the unfiltered spectrum is due to uncorrelated noise photons.
	}
	\label{fig:marg_spec}
\end{figure}

We can subtract the noise photons from the $g^{(2)}$ measurements in order to get a better comparison with $K$.
Following Ref.~\cite{eckstein2011realistic}, the noise-subtracted $\tilde{g}^{(2)}$ is given by:
\begin{equation}
    \tilde{g}^{(2)} = \frac{(1+R-Rp)^2 g^{(2)} - R^2}{1-p^2R^2}
\end{equation}
where $p\sim0.01$ is the probability per pulse to generate a photon and $R$ is the fraction of noise photons.
We estimate $R$ by comparing the spectra of the down-converted modes measured with and without the bandpass filters.
The signal's spectrum is shown in Fig.~\ref{fig:marg_spec}.
A flat background is apparent in the unfiltered spectrum.
By drawing a straight line at the background level [dashed line], we can estimate the fraction of the area under the blue curve occupied by noise photons [grey box].
We find that the noise photons contribute to approximately $R \sim 4(2)\%$ of the total counts with the filters in place. 
The uncertainty in $R$ arises from the human error in placing the dashed line and the edges of the box.

%% UNCOMMENT FOR SEPERATE DOCUMENT

% \bibliographystyle{apsrev4-1}
% \bibliography{refs}

% \end{document}